# An Empirical Study on variants of TCP over AODV routing protocol in MANET

Md. Monzur Morshed* [1, 2], Meftah Ur Rahman[2, 3], Md. Rafiqul Islam[1]
{m.monzur@gmail.com, monzur@tigerhats.org},
tasnimbd@gmail.com,
rafiqulislam@aiub.edu
Department of Computer Science
American International University-Bangladesh[1]
TigerHATS Research Team-Bangladesh[2]
George Mason University-USA[3]

*Abstract*—The cardinal concept of TCP development was to carry data within the network where network congestion plays a vital role to cause packet loss. On the other hand, there are several other reasons to lose packets in Mobile Ad Hoc Networks due to fading, interfaces, multi-path routing, malicious node, and black hole. Along with throughput, fairness of TCP protocols is important to establish a good communication. In this paper, an empirical study has been done by simulation and analysis of TCP variations under AODV routing protocol. In our simulation, we studied multiple variations of TCP, such as Reno, New-Reno, Vegas, and Tahoe. The simulation work has been done in NS2 environment. Based on the analysis simulation result of we carried out our observations with respect to the behavior of AODV routing protocol for different TCP packets under several QoS metrics such as drop, throughput, delay, and jitter.

## I. INTRODUCTION

TCP is one of the most popular end-to-end protocols offers reliable connection and compatible for both in wired and wireless networks. It was originally developed for wired networks with the mechanism of keeping low Bit Error Rate (BER). Later, the idea of forming an ad hoc on-the-fly network of mobile devices opens up an exciting new world of possibilities. Because ad-hoc networks do not need any preconfigured infrastructure, they can solve many interesting problems of spontaneous link establishment, *i.e.* communication on the fly. In this case, ad-hoc networks have a clear advantage over the classic, wire-bound connections. However, unlike wired links, wireless radio channels are affected by many factors that may lead to high levels of BER [1]. Though, TCP does not have the functionality to determine the packet loss where the reasons can be network congestion, channel errors, link failure, fading, interfaces, multi-path routing, malicious nodes and black hole etc., it has been the dominant transport-layer protocol providing reliable byte stream delivery between end-host applications with mechanism of connection management, congestion control, flow control, and error control [2].

MANET or mobile Ad hoc network is a collection of mobile nodes that are dynamically and arbitrarily located in such a manner that the interconnections between nodes are capable of changing on a continual basis. In order to facilitate communication within the network, a routing protocol is used to discover routes between nodes. After path establishment, either connection oriented protocol such as TCP or connection less protocol *i.e.* UDP is necessary to transfer the actual data packets. Due to its reliability, TCP and its variants play a crucial role in data transfer over MANET.

## II. AD-HOC ON-DEMAND DISTANCE VECTOR (AODV)

AODV routing protocol is an on demand routing protocol. To find a route to the destination, the source node floods the network with RouteRequest packets. The RouteRequest packets create temporary route entries for the reverse path through every node it passes in the network. When it reaches the destination a RouteReply is sent back through the same path the RouteRequest was transmitted. Every node maintains a route table entry which updates the route expiry time. A route is valid for the given expiry time, after which the route entry is deleted from the routing table. Whenever a route is used to forward the data packet the route expiry time is updated to the current time plus the Active Route Timeout. An active neighbor node list is used by AODV at each node as a route entry to keep track of the neighboring nodes that are using the entry to route data packets. These nodes are notified with RouteError packets when the link to the next hop node is broken. Each such neighbor node, in turn, forwards the RouteError to its own list of active neighbors, thus invalidating all the routes using the broken link [3, 4, 5].

## III. TCP TAHOE

Tahoe refers to the TCP congestion control algorithm which was suggested by Van Jacobson in his paper [6]. This implementation added a number of new algorithms and refinements to earlier implementations. The new algorithms include Slow-Start, Congestion Avoidance, and Fast Retransmit. TCP is based on a principle of 'conservation of packets', i.e. if the connection is running at the available bandwidth capacity then a packet is not injected into the network unless a packet is taken out as well. TCP implements this principle by using the acknowledgements to clock outgoing packets because an acknowledgement means that a packet was taken off the wire by the receiver. It also maintains a congestion window CWD to reflect the network capacity [6].

This research is supported by AIUB & TigerHATS Research Team.
For more information please visit www.aiub.edu, www.tigerhats.org

However there are certain issues, which need to be resolved to ensure this equilibrium:

i) Determination of the available bandwidth.
ii) Ensuring that equilibrium is maintained.
iii) How to react to congestion.

*A. Slow Start*

TCP packet transmissions are clocked by the incoming acknowledgements. However there is a problem- when a connection first starts up it needs to have acknowledgements so we need to have data in the network and to put data in the network we again need acknowledgements. To get around this circularity, Tahoe suggests that whenever a TCP connection starts or re-starts after a packet loss it should go through a procedure called 'slow-start'. The reason for this procedure is that an initial burst might overwhelm the network and the connection might never get started. Slow start suggests that the sender set the congestion window to 1 and then for each ACK received it increase the window by 1. So in the first round trip time (RTT) we send 1 packet, in the second we send 2 and in the third we send 4. Thus we increase exponentially until we lose a packet which is a sign of congestion. When we encounter congestion we decrease our sending rate and reduce congestion window to one and start over again. The important thing is that Tahoe detects packet losses by timeouts. In usual implementations, repeated interrupts are expensive so we have coarse grain time-outs which occasionally checks for time outs. Thus it might be some time before we notice a packet loss and then re-transmit that packet.

*B. Congestion Avoidance*

For congestion avoidance, Tahoe uses 'Additive Increase Multiplicative Decrease'. A packet loss is taken as a sign of congestion and Tahoe saves the half of the current window as a threshold value. It then set the congestion window to one and starts slow start until it reaches the threshold value. After that it increments linearly until it encounters a packet loss. Thus it increases its window slowly as it approaches the bandwidth capacity.

## IV.   TCP RENO

TCP Reno is the most widely adopted Internet TCP protocol. It retains the basic principle of Tahoe, such as slow starts and the coarse grain re-transmit timer. However it adds some intelligence over it so that lost packets are detected earlier and the pipeline is not emptied every time a packet is lost. It employs four transmission phases: slow start, congestion avoidance, fast retransmit, and fast recovery. When packet loss occurs in a congested link due to buffer overflow in the intermediate routers, either the sender receives three duplicate acknowledgments or the sender's retransmission timeout (RTO) timer expires. Thus, TCP Reno requires that we receive immediate acknowledgement whenever a segment is received. The logic behind this is that whenever we receive a duplicate acknowledgment, then his duplicate acknowledgment could have been received if the next segment in sequence expected, has been delayed in the network and the segments reached there out of order or else that the packet is lost. If we receive a number of duplicate acknowledgements then that means that sufficient time have passed and even if the segment had taken a longer path, it should have gotten to the receiver by now. There is a very high probability that it was lost. So Reno suggests an algorithm called 'Fast Re- Transmit'. Whenever we receive 3 duplicate acknowledgements, we take it as a sign that the segment was lost, so we re-transmit the segment without waiting for a timeout. Thus we manage to re-transmit the segment with the pipe almost full. Another modification that RENO makes is in that after a packet lost, it does not reduce the congestion window to 1. Since this empties the pipe. It enters into an algorithm which we call 'Fast-Re-Transmit' [7, 8].

## V.   NEW-RENO

New RENO is a slight modification over TCP-RENO. It is able to detect multiple packet losses and thus is much more efficient that RENO in the event of multiple packet losses. Like Reno, New-Reno also enters into fast-retransmit when it receives multiple duplicate packets; however it differs from RENO in that it doesn't exit fast-recovery until all the data which was out standing at the time it entered fast-recovery is acknowledged. Thus it overcomes the problem faced by Reno of reducing the congestion window size multiples times. The fast-transmit phase is the same as in Reno. The difference is the fast-recovery phase which allows for multiple re-transmissions in new-Reno [9]. TCP New-Reno exits fast recovery after receiving acknowledgement of all unacknowledged segments. It then sets congestion window size to slow start threshold and continues the congestion avoidance phase [10]. It retransmits the next segment when it receives a partial acknowledgment. (Partial acknowledgments are the acknowledgments that do not acknowledge all outstanding packets at the onset of the fast recovery.)

## VI.   VEGAS

Vegas is a TCP implementation which is a modification of Reno. It builds on the fact that proactive measures to encounter congestion are much more efficient than reactive ones. It tried to get around the problem of coarse grain timeouts by suggesting an algorithm which checks for timeouts at a very efficient schedule. Also it overcomes the problem of requiring enough duplicate acknowledgements to detect a packet loss, and it also suggests a modified slow start algorithm which prevents it from congesting the network. It does not depend solely on packet loss as a sign of congestion. It detects congestion before the packet losses occur. However it still retains the other mechanism of Reno and Tahoe, and a packet loss can still be detected by the coarse grain timeout of the other mechanisms fail. The three major changes induced by Vegas are:

## A. New Re-Transmission Mechanism

Vegas extend on the re-transmission mechanism of Reno. It keeps track of when each segment was sent and it also calculates an estimate of the RTT by keeping track of how long it takes for the acknowledgment to get back. Whenever a duplicate acknowledgement is received it checks to see if the (current time - segment transmission time) > RTT estimate or not. If it is, then it immediately retransmits the segment without waiting for 3 duplicate acknowledgements or a coarse timeout. Thus it gets around the problem faced by Reno of not being able to detect lost packets when it had a small window and it didn't receive enough duplicate acknowledgements. To catch any other segments that may have been lost prior to the retransmission, when a non duplicate acknowledgment is received, if it is the first or second one after a fresh acknowledgement then it again checks the timeout values and if the segment time since it was sent exceeds the timeout value then it re-transmits the segment without waiting for a duplicate acknowledgment. Thus in this way Vegas can detect multiple packet losses. Also it only reduces its window if the re-transmitted segment was sent after the last decrease. Thus it also overcomes Reno's shortcoming of reducing the congestion window multiple time when multiple packets are lost.

## B. Congestion Avoidance

TCP Vegas is different from all the other implementation in its behavior during congestion avoidance. It does not use the loss of segment to signal that there is congestion. It determines congestion by a decrease in sending rate as compared to the expected rate, as result of large queues building up in the routers. It uses a variation of Wang and Crowcroft's Tri-S scheme. Thus whenever the calculated rate is too far away from the expected rate it increases transmissions to make use of the available bandwidth, whenever the calculated rate comes too close to the expected value it decreases its transmission to prevent over saturating the bandwidth. Thus Vegas combats congestion quite effectively and doesn't waste bandwidth by transmitting at too high a data rate and creating congestion and then cutting back, which the other algorithms do.

## C. Modified Slow-start

TCP Vegas differs from the other algorithms during its slow-start phase. The reason for this modification is that when a connection first starts it has no idea of the available bandwidth and it is possible that during exponential increase it over shoots the bandwidth by a big amount and thus induces congestion. To this end Vegas increases exponentially only every other RTT, between that it calculates the actual sending through put to the expected and when the difference goes above a certain threshold it exits slow start and enters the congestion avoidance phase [11].

## VII. SIMULATION TOPOLOGY

There Simulation environment consists of 16 wireless mobile nodes which are placed uniformly and forming a Mobile Ad-hoc Network, moving about over a 1000 × 1000 meters area for 40 seconds of simulated time. We have used standard two-ray ground propagation model, the IEEE 802.11 MAC, and Omni-directional antenna model of NS2. We have used AODV routing algorithm and interface queue length 50 at each node. The source nodes are respectively 6, 15 and 5 and the receiving nodes are respectively 0, 1 and 11.

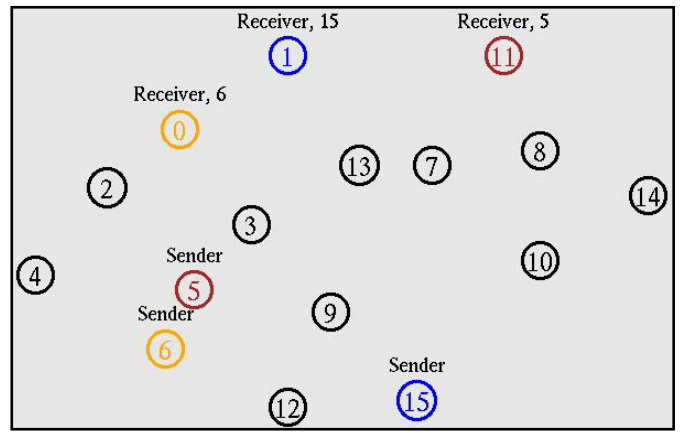

Figure 1. Simulation Topology in NS2 environment

## VIII. SIMULATION DESCRIPTION

TABLE I. SIMULATION PARAMETERS

| Method | Value |
|---|---|
| Channel type | Channel/Wireless channel |
| Radio-propagation model | Propagation/Two ray round |
| Network interface type | Phy/wirelessphy |
| MAC type | Mac/802.11 |
| Interface queue type | Queue/Drop Tail |
| Link Layer Type | LL |
| Antenna | Antenna/omni antenna |
| Maximum packet in ifq | 50 |
| Area (m×m) | 1000×1000 |
| Number of mobile nodes | 16 |
| Source type | TCP (Tahoe, Reno, NewReno, Vegas) |
| Simulation Time | 40 sec |
| Routing protocol | AODV |

## IX. QOS METRICS

We used different parameter of QoS metrics such as delay, jitter, packet drop and throughput to understand the behavior of TCP in AODV Routing Protocol.

## X. SIMULATION RESULT

### A. Drop

The routers might fail to deliver (drop) some packets if they arrive when their buffers are already full. Some, none, or all of the packets might be dropped, depending on the state of the network, and it is impossible to determine what will happen in advance. The receiving application may ask for this information to be retransmitted, possibly causing severe delays in the overall transmission. Packet drop is equal to number of packets sent from source minus number of packet received in the path of destination *i.e.*

*No of Packets Dropped = No of pkt Sent – No of pkt Received*

TABLE II.   NUMBERS OF PACKETS DROP

| Packets | Total Sent packets | Total Received packets | Total dropped packets |
|---|---|---|---|
| Tahoe | 626 | 580 | 11 |
| Reno | 593 | 530 | 25 |
| New Reno | 615 | 550 | 29 |
| Vegas | 503 | 489 | 8 |

### B. Throughput

Throughput is the measurement of number of packets passing through the network in a unit of time. This metric show the total number of packets that have been successfully delivered to the destination nodes and throughput improves with increasing nodes density. Throughput can be defined by:

$$\frac{\Sigma \text{ Node Throughputs of Data Transmission}}{\text{Total number of nodes}}$$

TABLE III.   DATA FOR THROUGHPUT

| Packets | Total Sending Throughput | Total Receiving Throughput |
|---|---|---|
| Tahoe | 501248 | 438368 |
| Reno | 501248 | 453796 |
| New Reno | 501248 | 445816 |
| Vegas | 501248 | 462840 |

### C. Delay

A specific packet is transmitting from source to destination and calculates the difference between send times and received times. Delays due to route discovery, queuing, propagation and transfer time are included in the delay metric.

*Packet Delay = packets receive time – packet send time*

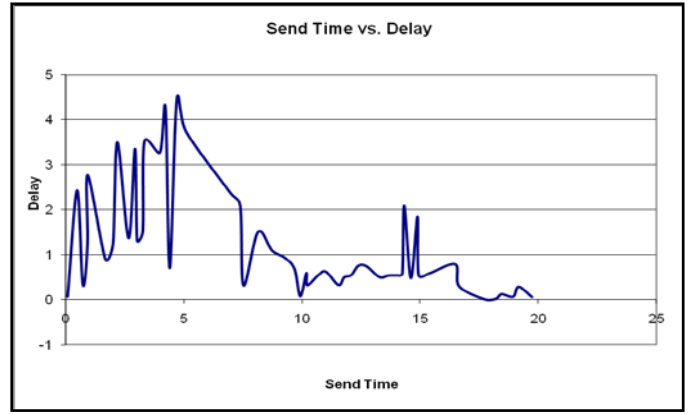

Figure 2. Delay for TCP Tahoe

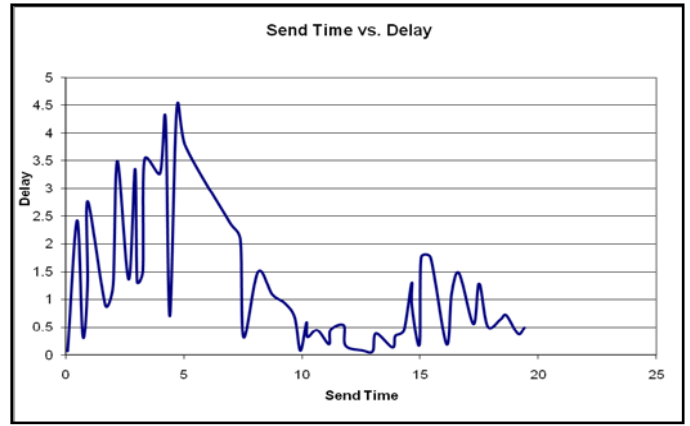

Figure 3. Delay for TCP Reno

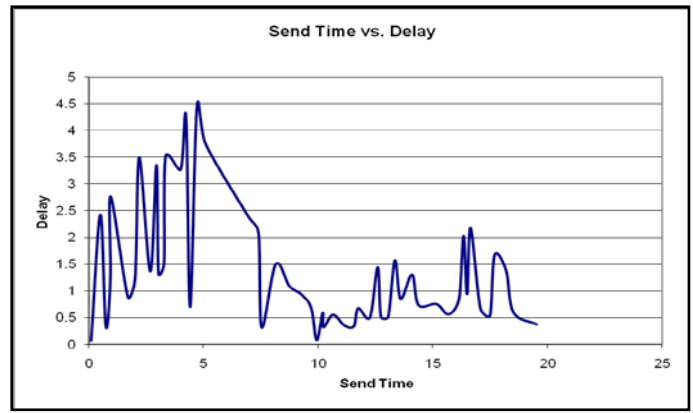

Figure 4. Delay for TCP NewReno

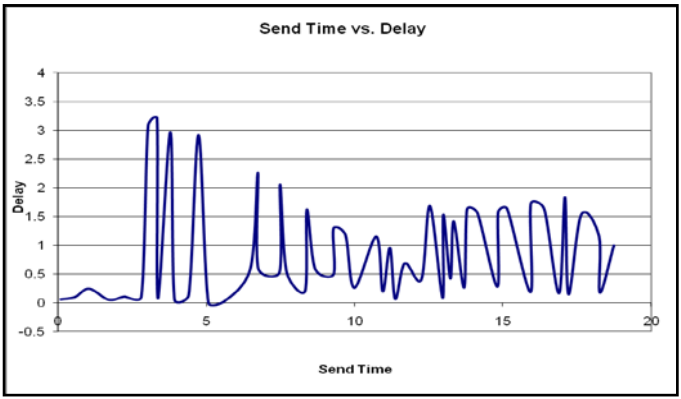

Figure 5. Delay for TCP Vegas

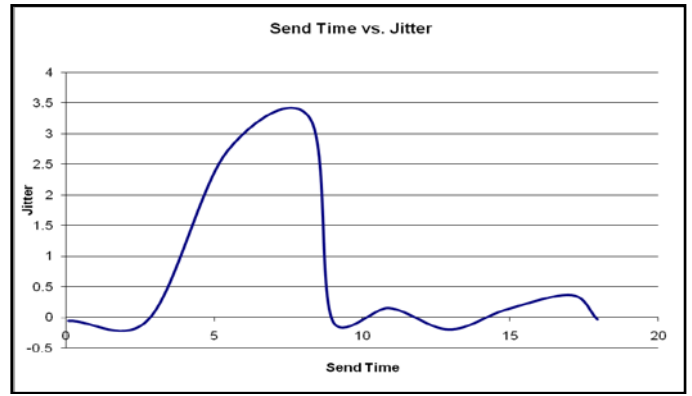

Figure 8. Jitter for TCP NewReno

*D. Jitter*

Jitter is the variation of the packet arrival time. In jitter calculation the variation in the packet arrival time is expected to minimum. The delays between the different packets need to be low if we want better performance in Mobile Ad-hoc Networks.

*Jitter ( i ) = Delay (i+1) – Delay (i)  where i =  1,2,3.....n*

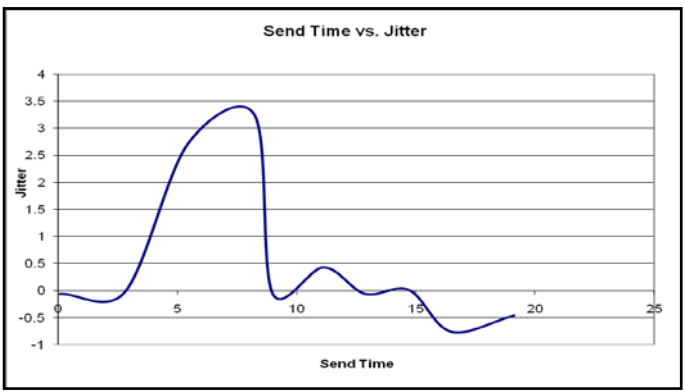

Figure 6. Jitter for TCP Tahoe

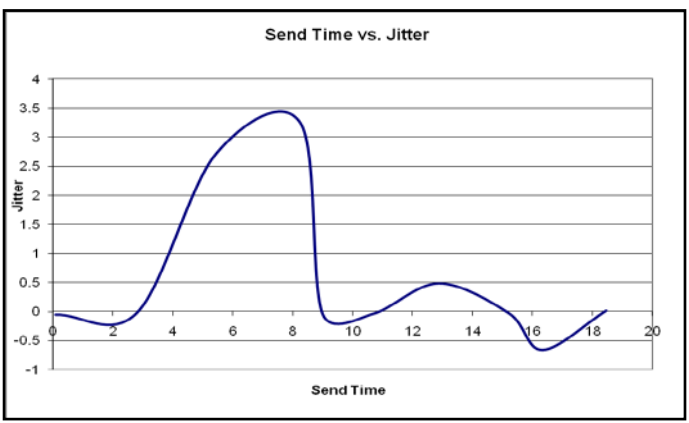

Figure 7. Jitter for TCP Reno

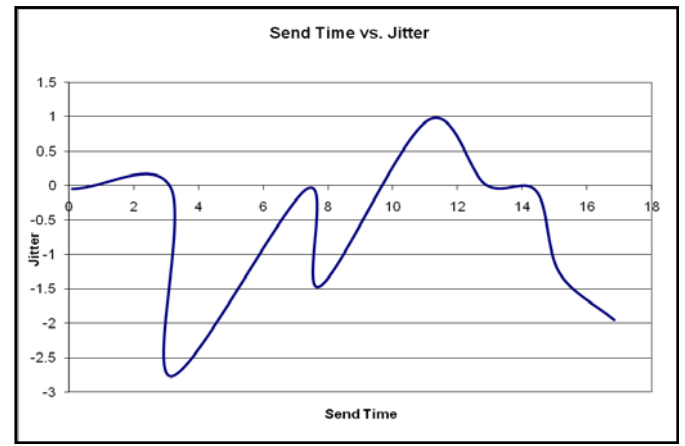

Figure 9. Jitter for TCP Vegas

## XI.  CONCLUSION

We carried out our simulation work for four types of TCP variants and analyzed TCP variants where the packet drop rates for Tahoe, Reno, NewReno and Vegas are respectively 1.75%, 4.21%, 4.71%, 1.59%. For TCP routing variants, the packet delivery ratio is independent of offered traffic load and are respectively 92.65%, 89.38%, 89.43% and 97.22% for each one. So we can conclude that considering the performance on the variants of TCP, Vegas proves to be showing the highest efficiency and perform best. In terms of drop rates and throughput, Vegas is clearly best among the four variants. However, similar decision can not be reached when we consider Jitter and Delay as QoS metrices. We can see that over time, both of these matrices show similar patterns of change for Tahoe, Reno and NewReno. But for Vegas, the deviation of graph is somewhat random. It can be assumed that further research is necessary to properly understand the efficacy of using different TCP variants in Mobile Ad Hoc Networks.